\documentclass[prb,twocolumn,showpacs,nofootinbib,
superscriptaddress,preprintnumbers,amsmath,amssymb]{revtex4}
\usepackage{graphicx}
\usepackage{amsbsy}
\usepackage{dcolumn}
\usepackage{bm}
\begin{document}
\title{Search for broken time-reversal symmetry near 
  the surface of \YBCO\ films using $\beta$-NMR}
\author{H. Saadaoui}
\altaffiliation[Present address: ]{Paul Scherrer Institute, Laboratory
for Muon Spin Spectroscopy, 5232 Villigen PSI, Switzerland.}
\affiliation{Department of Physics and Astronomy, University of British
  Columbia, Vancouver, BC, V6T 1Z1, Canada}
\author{G.~D. Morris}
\affiliation{TRIUMF, 4004 Wesbrook Mall, Vancouver, BC, V6T 2A3, Canada}
\author{Z. Salman}
\altaffiliation[Present address: ]{Paul Scherrer Institute, Laboratory
for Muon Spin Spectroscopy, 5232 Villigen PSI, Switzerland.}
\affiliation{Clarendon Laboratory, Department of Physics,
  Oxford University, Parks Road, Oxford, OX1 3PU, UK}
\author{Q. Song}
\affiliation{Department of Physics and Astronomy, 
  University of British Columbia, Vancouver, BC, V6T 1Z1, Canada}
\author{K.~H. Chow}
\affiliation{Department of Physics, University of Alberta, 
  Edmonton, AB, T6G 2G7, Canada}
\author{M.~D. Hossain}
\affiliation{Department of Physics and Astronomy, 
  University of British Columbia, Vancouver, BC, V6T 1Z1, Canada}
\author{C. D. P. Levy}
\affiliation{TRIUMF, 4004 Wesbrook Mall, Vancouver, BC, V6T 2A3, Canada}
\author{T.~J. Parolin}
\affiliation{Chemistry Department, University of British Columbia, Vancouver,
  BC, V6T 1Z1, Canada}
\author{M. R. Pearson}
\affiliation{TRIUMF, 4004 Wesbrook Mall, Vancouver, BC, V6T 2A3, Canada}
\author{M. Smadella}
\affiliation{Department of Physics and Astronomy, 
  University of British Columbia, Vancouver, BC, V6T 1Z1, Canada}
\author{D. Wang}
\affiliation{Department of Physics and Astronomy, 
  University of British Columbia, Vancouver, BC, V6T 1Z1, Canada}
\author{L.~H. Greene}
\affiliation{Department of Physics, 
  University of Illinois at Urbana-Champaign, Urbana, Illinois, 61801, USA}
\author{P.~J. Hentges}
\affiliation{Department of Physics, 
  University of Illinois at Urbana-Champaign, Urbana, Illinois, 61801, USA}
\author{R.~F. Kiefl}
\affiliation{Department of Physics and Astronomy, 
  University of British Columbia, Vancouver, BC, V6T 1Z1, Canada}
\affiliation{TRIUMF, 4004 Wesbrook Mall, Vancouver, BC, V6T 2A3, Canada}
\affiliation{Canadian Institute for Advanced Research, 
  Toronto, ON, M5G 1Z8, Canada}
\author{W.~A. MacFarlane}
\affiliation{Chemistry Department, 
  University of British Columbia, Vancouver, BC, V6T 1Z1, Canada}
\date{\today}

\newcommand{\LSCO}{La$_{2-x}$Sr$_x$CuO$_{4-\delta}$}
\newcommand{\YBCO}{YBa$_{2}$Cu$_{3}$O$_{7-\delta}$}
\newcommand{\BISCO}{Bi$_2$Sr$_2$Cu$_2$O$_{8+\delta}$}
\newcommand{\STO}{SrTiO$_{3}$}
\newcommand{\SRO}{Sr$_2$RuO$_{4}$}
\newcommand{\PCCO}{Pr$_{2-x}$Ce$_x$CuO$_{4-\delta}$}
\newcommand{\CO}{$\rm CuO_2$}
\newcommand{\Li}{${}^8$Li$^+$}
\newcommand{\Lip}{${}^8$Li$^+$}
\newcommand{\NbSe}{NbSe$_2$}

\newcommand{\msr}{$\mu$SR}
\newcommand{\lem}{LE$\mu$SR}
\newcommand{\bnmr}{$\beta$-NMR}
\newcommand{\spct}{superconductivity}
\newcommand{\Spct}{Superconductivity}
\newcommand{\spc}{superconductor}
\newcommand{\Spc}{Superconductor}
\newcommand{\etal}{{\it et al.}}
\newcommand{\Htsc}{High-$T_C$ superconductors}
\newcommand{\htsc}{high-$T_C$ superconductors}
\newcommand{\ie}{{\it i.e.}}
\newcommand{\Tc}{$T_c$}

\newcommand{\PRL}[3]{Phys. Rev. Lett. {\bf #1}, {#2} ({#3})}
\newcommand{\PRB}[3]{Phys. Rev. B {\bf {#1}}, {#2} ({#3})}
\newcommand{\PB}[3]{Physica B {\bf {#1}}, {#2} ({#3})}
\newcommand{\PC}[3]{Physica C {\bf {#1}}, {#2} ({#3})}
\newcommand{\Nt}[3]{Nature {\bf {#1}}, {#2} ({#3})}
\newcommand{\Sc}[3]{Science {\bf {#1}}, {#2} ({#3})}
\newcommand{\RMP}[3]{Rev. Mod. Phys. {\bf {#1}}, {#2} ({#3})}
\newcommand{\JPSJ}[3]{J. Phys. Soc. Jap. {\bf #1}, {#2} ({#3})}
\newcommand{\RPP}[3]{Rep. Prog. Phys. {\bf {#1}}, {#2} ({#3})} 
\newcommand{\ibid}[3]{{\it ibid}. {\bf {#1}}, {#2} ({#3})} 
 \newcommand{\equ}[2]{\begin{equation}\label{#1}{#2}\end{equation}}
\newcommand{\meq}[2]{\begin{eqnarray}\label{#1}{#2}\end{eqnarray}}
\newcommand{\bq}{{\bf q}}
\newcommand{\bk}{{\bf k}}
\newcommand{\bkp}{{\bf k'}}
\newcommand{\br}{{\bf r}}
\newcommand{\bR}{\bf R}
\newcommand{\bp}{{\bf p}}

\begin{abstract}
  Weak spontaneous magnetic fields are 
  observed near the surface of \YBCO\ films 
  using $\beta$-detected Nuclear Magnetic Resonance. Below $T_c$,
  the magnetic field distribution in a silver film 
  evaporated onto the superconductor shows additional line broadening,
  indicating the appearance of small disordered magnetic fields. 
  The line broadening increases linearly 
  with a weak external magnetic field applied parallel to the surface, 
  and is depth-independent up to 45 nm from the Ag/\YBCO\ interface.
  The  magnitude of the line broadening  extrapolated 
   to zero applied field is less than 0.2 G, and is close 
  to  nuclear dipolar broadening in the Ag. This indicates that any 
  fields due to  broken time-reversal symmetry are less than 0.2 G.
\end{abstract}
\maketitle
The highly unconventional electronic properties of 
high-\Tc\ superconductors (HTSC) give rise to novel interfacial 
phenomena that are important fundamentally
(e.g. in probing symmetry of the bulk electronic ground 
state); as well as in applications (e.g. junction-based
devices). While
significant progress has been made in understanding the transport properties
of such interfaces, very little is known about their {\it magnetic}
properties, in part due to the lack of an appropriate local magnetic probe. 
A particularly unresolved issue is whether the superconducting order
parameter (OP) breaks time-reversal symmetry (TRS) 
near the surface.\cite{Gorkov87,SigristRMP91}
A characteristic feature of TRS-breaking (TRSB) 
is spontaneous magnetization; however,
Meissner screening cancels this in the bulk,
limiting the associated fields to within the magnetic penetration depth
of  defects and interfaces.\cite{sigristPTP98}
To measure this  magnetization directly, one requires 
a sensitive depth-dependent local magnetic probe.
In this paper we use a novel technique based on depth-controlled 
beta-detected  nuclear magnetic resonance (\bnmr)
to search for TRSB order 
near the surface of the high-$T_c$ cuprate superconductor \YBCO\ (YBCO).

In contrast to Ru-based and heavy fermion 
superconductors,\cite{LukeNt98,LukePRL93} 
there is no evidence for TRSB in the bulk of HTSC cuprates, 
particularly YBCO,\cite{kieflPRL90} where OP-phase-sensitive
measurements have established spin-singlet 
$d_{\rm x^2-y^2}$-wave order.\cite{vanHarlingenRMP95} 
There are some indications of weak magnetism,\cite{sonierSc01,xiaPRL08}
some of it related to the CuO chains in YBCO,\cite{yamaniPRB06}
or to vortex cores above the lower critical field $H_{c1}$.\cite{sonierRPP07}
Surface scattering of the Cooper pairs 
from most surfaces perpendicular to the ${\rm CuO}_2$ 
planes frustrates $d_{\rm x^2-y^2}$-wave
order within a few coherence lengths of the interface.\cite{sigristPTP98} 
This leaves a high density of mobile holes
(as evidenced by the zero bias conductance peak (ZBCP)
in tunneling spectra) that may condense into
a superfluid of different symmetry than the bulk,\cite{huPRL94} e.g.
$s$-wave, or TRSB states such as $d_{x^2-y^2}+is$
and $d_{x^2-y^2}+id_{xy}$.\cite{fogelstromPRL97}
Other origins of a TRSB state include, frustrated OP near grain boundaries or
junctions,\cite{sigristPTP98} the  
interaction of a self-induced magnetic field caused 
by the OP distortion with the OP itself,\cite{palumboPRB92}
and finite size effects in thin films.\cite{VorontsovPRL09,AminPRB02}

Experiments to detect TRSB near surfaces have yielded controversial results. 
Carmi {\it et al.} measured a weak spontaneous magnetic field 
using SQUID magnetometry near the edges of 
epitaxial c-axis oriented YBCO films below \Tc.\cite{carmiNt00}
Spontaneous Zeeman-like splitting of the ZBCP, due to 
TRSB, has been seen in some tunneling 
measurements,\cite{covingtonPRL97} but not in others.\cite{deutcherRMP05} 
Phase sensitive measurements, which could also detect spontaneous flux,
showed no evidence of a TRSB state.\cite{neilsPRL02} 
A resolution to this disagreement requires more direct information 
of interface magnetism in cuprates using a local magnetic 
probe that can locate the origin and distribution of any such fields 
on the atomic scale.

In this study, we present direct measurements of the magnetic field near 
the interface of silver and YBCO films using $\beta$-NMR. 
We measure the field distribution  using  a highly spin 
polarized \Li\ beam implanted
into a thin silver overlayer deposited on YBCO.
We find an inhomogeneous broadening of the 
field distribution below the $T_c$ of YBCO,
with the probe ions stopping at an average distance 
of 8 nm from the Ag/YBCO interface.
By extrapolating the line  broadening to zero applied field 
we find  the mean internal field is very close to experimental 
resolution determined by  the Ag nuclear dipolar fields. 
In this way we obtain an upper limit of 0.2 G for any spontaneous fields  
of electronic origin.

The experiment was performed using \bnmr\ of  
\Li\ at the ISAC facility at TRIUMF in Vancouver, Canada.
For details, see Refs.\cite{kieflPC03,morrisPRL04}
Similar to NMR, to measure the resonance, 
we apply a field along the
spin polarization (here in the plane of the films) 
${\bf B}_0=B_0\hat{y}$ (with $5\le{B_0}\le150$ G)
and follow the polarization of \Li\ as a function of the frequency $\omega$ of 
a small transverse radio-frequency (RF) field of amplitude 
$B_1\sim 1$ G, applied along the $\hat{x}$-axis. 
The resonance condition is $\omega = \gamma B_{\rm loc}$,
where for \Li, $\gamma=0.63015$ kHz/G, and 
$B_{\rm loc}$ is the local field.
At this $\omega$, the polarization, initially parallel to the $\hat{y}$-axis,
is averaged by precession in the oscillating field.
In the absence of dynamic effects, 
the resulting resonance is generally broadened
by any static inhomogeneity in the local magnetic field.
Thus, the lineshape offers a detailed measurement
of the distribution of local magnetic fields 
in the sampled volume determined by
the beamspot ($\sim 3$ mm in diameter) 
and the implantation profile (see below). 

\begin{figure}
  \includegraphics[width=\columnwidth]{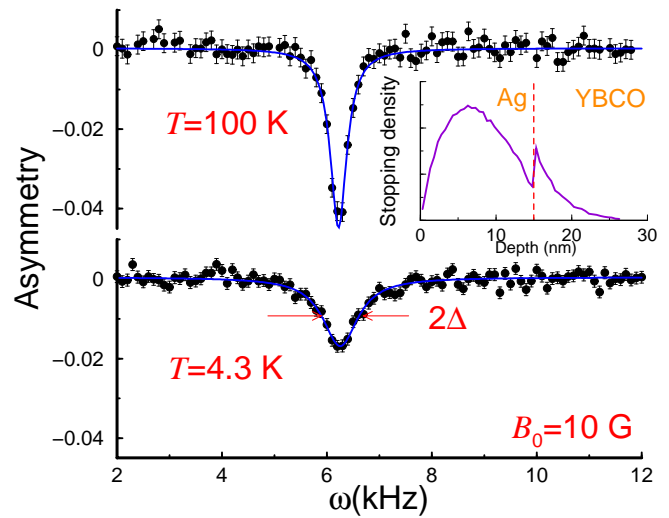}
  \caption{(Color online). Typical \bnmr\ spectra  taken 
    by implanting 2 keV \Li\ into Ag/YBCO(110),  
    in an external field of $B_0=10$ G (FC) applied along the
    surface of the film.  Solid lines are fits to a Lorentzian of 
    HWHM $\Delta$. Inset: simulated implantation profile using TRIM.SP for 
    \Li\ of 2 keV in 15 nm of Ag on YBCO.\cite{trim.sp}
    The \Li\ stops at an average depth 
    of 8 nm away from the Ag/YBCO interface. }
  \label{spectra}
\end{figure}

A novel pulsed RF mode was used in this study. The RF field
is applied in $90^{\rm o}$ pulses randomized in frequency order, 
instead of the continuous wave (cw) mode commonly used.\cite{morrisPRL04}
In randomly pulsed RF (RPRF), one obtains a high signal 
to noise ratio with minimal contribution from both
variations in the incoming \Li\ rate
and cw power broadening. Because of the limited $B_1$,
the RPRF mode is suitable for narrow lines up to a few kHz in width.
Fig. \ref{spectra} shows the resonance 
spectrum at 100 K with a half width at half maximum (HWHM)  
of approximately $\Delta_0\approx 0.15$ kHz (or 0.24 G). 
Similar cw spectra can be at least twice as broad,\cite{kieflPC03} 
making it difficult to resolve a small additional 
broadening.

Major advantages of \bnmr\ in detecting TRSB are the abilities 
(i) to implant the probe \Li\ at low energy into thin layered structures and 
(ii) to control the mean implantation depth on the nanometer scale.
In this study, \Li\ is preferentially implanted into the thin silver
overlayer evaporated onto YBCO, instead of the superconductor itself.
Stopping the probes in the overlayer eliminates the possibility 
of the probe perturbing the superconductor. Also, the \Li\ nucleus carries 
a small electric quadrupole moment, so the spectrum in 
the Ag is free  of any quadrupole splittings 
which are present in non cubic YBCO.\cite{kieflPC03}
Consequently the resonance of \Li\ in Ag, below 1 Tesla, is  
single narrow line with a a $T$-independent linewidth  
attributed to  nuclear dipole broadening from the small 
nuclear moments of ${}^{107}$Ag and ${}^{109}$Ag.\cite{morrisPRL04}
From basic magnetostatics, any inhomogeneous fields 
in the YBCO layer will decay 
exponentially outside the superconductor as $\exp(-\frac{2\pi}{a}z)$, where $a$
is the length scale of the inhomogeneity.\cite{xuJMR07,salmanNl07,BluhmPRB07}
Thus we can only detect such fields provided 
our probe-YBCO stopping distance $z$
is $\lesssim \frac{a}{2\pi}$. Any static field inhomogeneities
arising in this way will broaden the intrinsic resonance of the Ag layer.

The measurements presented here were carried out on (110), 
(103), and (001)-oriented  
YBCO films capped with 15 or 50 nm of Ag.
The (110) film ($T_c=86.7$ K) was grown by
RF magnetron sputtering on a (110) SrTiO$_3$ (STO) 
substrate measuring $8\times6$ mm.
The (103) film ($T_c=84$ K) was grown under similar conditions.
Three (001) films were also studied, (i) one with $T_c=88.7$ K 
grown on (001) STO 
under similar conditions as the (110), and (ii) two films ($T_c\sim 88.0$ K)
grown by thermal co-evaporation on $8\times10$ mm LaAlO$_3$.
The films are epitaxial and  atomic force microscopy was used to
  characterize the surface roughness which is in the range 6-12 nm. 
All samples were capped {\it ex-situ} with 15 nm of Ag, except
one of the last two which was capped with 50 nm, 
by DC sputtering 99.99$\%$ Ag at room 
temperature under  an Ar pressure of 30 
mtorr at a rate of 0.5 \AA/s while rotating the sample to ensure
uniformity. The \Li\ implantation energy was varied so that 
the probe ions are implanted at average depths ranging from 8 to 43 nm. 
The inset of Fig. \ref{spectra}, shows the simulated
stopping profile of 2 keV \Li\
ions in 15 nm of Ag on YBCO using TRIM.SP.\cite{trim.sp}
Here the average probe-YBCO distance is $\sim 8$ nm.
At 2 keV, about 20$\%$ of \Li\ ions stop in the YBCO, 
yielding no associated NMR signal due to
fast spin-lattice relaxation at low fields.
To measure the NMR resonance in Ag, a small field is required, and our
measurements were taken by field-cooling (FC) in a small field $B_0$ 
or zero field-cooling (ZFC). 
Residual magnetic fields were reduced to less than 30 mG 
normal to the surface when FC, or in all directions when ZFC.
The samples were aligned parallel to the field 
with an accuracy of at least $0.5^\circ$.
\begin{figure}
  \includegraphics[width=\columnwidth]{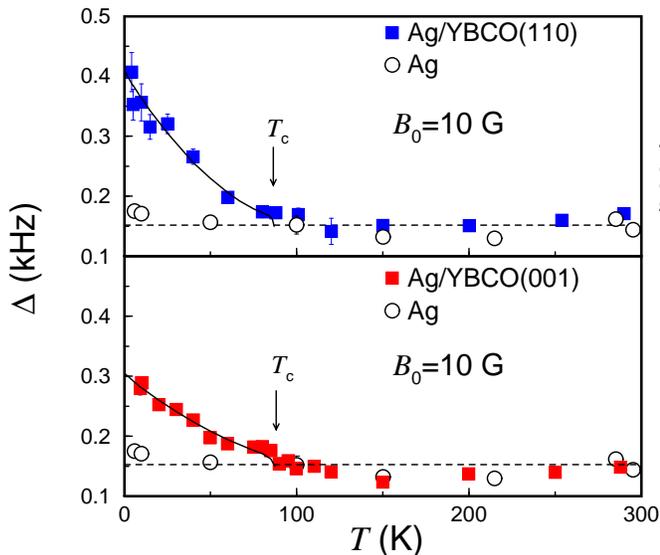}
  \caption{(Color online). The $T$-dependence of the HWHM, $\Delta$,  
    of the resonance 
    of 2 keV \Li\ implanted into Ag/YBCO(110), Ag/YBCO(001), and Ag. 
    The data on the Ag/YBCO(110) 
    and Ag film was taken using FC, and ZFC for the Ag/YBCO(001). 
    The widths were independent of FC or ZFC.
    The dashed lines represent the average Ag width $\Delta_0$, the arrows
    point to the $T_c$ of the YBCO films, 
    and the solid lines are guide to the eye. }
  \label{hwhmVsT}
\end{figure}

Fig. \ref{spectra} shows two resonances at 100 K and 4.3 K in Ag/YBCO(110). 
Above $T_c$, the resonances are all identical 
and show negligible differences in amplitude and linewidth, and
are indistinguishable from those intrinsic to Ag.
Below $T_c$, the resonance broadens, therefore is reduced in amplitude.
The broadening is symmetric, unlike the field distribution within the bulk of
a superconductor in an ordered vortex lattice state.\cite{sonierRMP}
Such a symmetric broadening is typical of a more disordered
vortex distribution.\cite{SaadaouiPRB10,MenonPRL06,HarshmanPRL91}
The HWHM, $\Delta$, of a single Lorentzian fit to both (110) and (001)
samples is plotted in Fig. \ref{hwhmVsT}. 
It is nearly $T$-independent  above $T_c$, consistent with the small  
nuclear dipolar broadening in Ag. It is the same in 
both  samples and comparable to a control sample of Ag grown on
an insulating STO substrate under similar conditions 
(open circles, Fig. \ref{hwhmVsT}). 
Below $T_c$, the resonance broadens, signaling the appearance of
disordered static magnetic fields in the underlying YBCO.

$\Delta$ below \Tc\ is slightly larger in Ag on the (110) film 
than the (001) film. In an applied field of 10 G, 
the additional broadening, $\Delta-\Delta_0$, 
at $\approx 5$ K is already very small, about 0.25 kHz for the (110) film,
and 0.15 kHz for the (001) film.  
This broadening is clearly caused by the superconducting YBCO, 
since it is absent above \Tc\ and in the Ag film without YBCO.
It is, however, not accompanied by a resonance
shift (see Fig. \ref{spectra}). The resonance frequency is
constant from 300 K to 5 K in all films, independent of FC or ZFC
cooling. This shows there is no superconducting 
proximity effect in the Ag layer
where an induced Meissner shielding of the applied field leads 
to a diamagnetic resonance shift, e.g. as seen recently in 
Ag/Nb heterostructures.\cite{morenzoni}
The temperature-dependent broadening below $T_c$ is not linked 
to TRSB, as the latter state
is expected theoretically to appear at a second transition temperature 
$T_{c2}\ll T_c$,\cite{huPRL94,sigristPTP98,fogelstromPRL97}
consistent with some tunneling measurements.\cite{covingtonPRL97}
In contrast, the broadening here has an onset close to $T_c$.
\begin{figure}
  \includegraphics[width=\columnwidth]{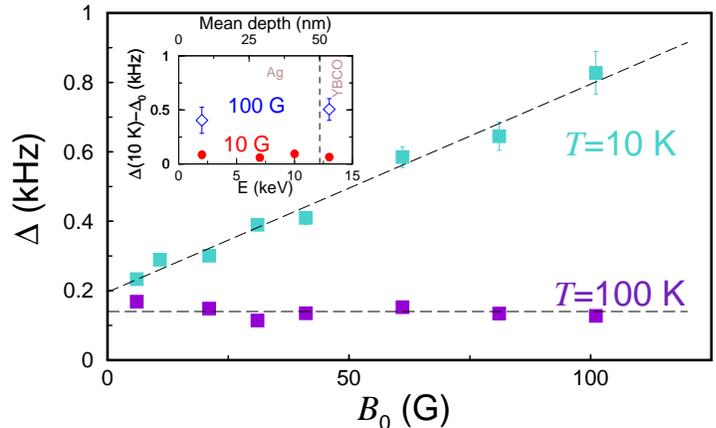}
  \caption{(Color online). The HWHM of the resonance 
    in Ag/YBCO(001) at 10 K taken
    after ZFC.
    The dashed lines are linear fits. 
    Inset: energy and depth dependence of the 
      excess HWHM at 10 K ($\Delta$(10 K)$-\Delta_0$) 
      in the 50 nm Ag/YBCO(001)
      sample at applied fields 10 and 100 G. 
      The resonance at all energies is due to the 
      fraction of \Li\ landing in Ag film.} 
  \label{hwhmVsB}
\end{figure}

The line broadening versus the external magnetic field in the c-axis sample 
is displayed in Fig. \ref{hwhmVsB}. At 100
K, $\Delta$ is field-independent as expected.  At 10 K (ZFC), $\Delta$
increases linearly with the applied field. 
This field-dependent broadening is attributed to  inhomogeneous
penetration of the applied field in the form of flux vortices.
Penetration of vortices would not typically occur 
at these fields well-below $H_{c1}$,\cite{ruixing94}
especially since the demagnetization factor for the field
parallel to the film is very small.
However, at the interface the flux lines may penetrate
more easily due to suppression of the $d$-wave order near 
twin or grain boundaries.\cite{AeppliPRB87,huebener} 
Moreover, surface roughness suppresses the Bean-Livingston surface barrier, 
and vortices may nucleate at fields $H\le H_{c1}$.\cite{konczykowskiPRB91}
Surface vortices have been observed in YBCO crystals
in fields as small as 4 G applied parallel to the surface.\cite{dolanPRL89} 
At such low fields, the vortex spacing 
$d$ is of the order of few microns.\cite{surface.vortices} 
Outside the superconductor,
the resulting field inhomogeneity leads to a depth independent broadening,
since $z \ll d$, consistent with our results (inset of Fig. \ref{hwhmVsB}).

The $T$-dependence of the linewidth in Fig. \ref{hwhmVsT} 
is consistent with  line broadening due to vortices  
beneath the surface since one would expect such vortex penetration 
close to $T_c$, and that it would increase  as  the temperature falls 
due to the decreasing magnetic  penetration depth in the 
superconductor.\cite{NiedermayerPRL99,PanagopoulosPRB99}
Field inhomogeneities are also expected from local variations of the shielding 
current density due to surface roughness
as well as twin and grain boundaries.\cite{joossPC96}
This inhomogeneity will be enhanced by increasing the field or 
decreasing the temperature below $T_c$; due to 
a higher current density with pronounced local variations.
This would lead to a temperature and field-dependent broadening as
observed in Figs. \ref{hwhmVsT} and \ref{hwhmVsB}. 
It is possible that flux penetration 
and current density inhomogeneities near the surface could be related to the 
apparent dead layer seen in HTSC and other
superconductors using low-energy \msr.\cite{KieflPRB10,SuterPRB05} 
 Further experiments on atomically flat surfaces 
may help elucidate the origin of the 
magnetic field inhomogeneities reported here.
However the field-dependent source of the line broadening is not  
central to the current study.

The main result of this study is the zero field extrapolation
of the broadening at low temperature which is an estimate of the TRSB fields.
The broadening at 10 K extrapolates to $\Delta_{B=0}\approx 0.2$ kHz in the
(001) and (103), and 0.3 kHz in (110) (not shown).  
This $\Delta_{B=0}$ is marginally higher than the normal state broadening, 
$\Delta_0 \approx 0.15$ kHz.
Thus, the net internal field in the superconducting state
extrapolated to zero applied field, $\Delta_{B=0}-\Delta_0$
is less than 0.15 kHz (or $\sim 0.2$ G) in all orientations.
Part of the difference is due to the slight broadening 
of the Ag resonance upon cooling from 100 to 5 K,
as seen in Fig. \ref{hwhmVsT} (open circles).
Thus, the additional broadening at zero field $\Delta_{B=0}-\Delta_0$,
is an estimate of the spontaneous magnetic field at the Ag/YBCO interface, 
and has an upper limit of 0.2 G. This extrapolated value is 
close to our experimental resolution determined by the Ag nuclear  moments.
These additional fields at low temperature
are clearly much weaker than predicted by tunneling experiments
where much stronger spontaneous fields are predicted  
to cause the ZBCP splitting.\cite{covingtonPRL97}

In conclusion, we have conducted a depth-resolved \bnmr\ study of the field
distribution near the interface of Ag and YBCO films.
In all films we find additional broadening of the NMR resonance below $T_c$,
signaling the appearance of disordered internal static fields in YBCO. 
We established an upper limit of 0.2 G for TRSB fields at low temperature.
This rules out any straightforward interpretation 
based on the TRSB state that was suggested by tunneling measurements
 to be characterized by a much larger
spontaneous magnetic field.\cite{covingtonPRL97,deutcherRMP05}
We have shown that such a putative state must be consistent with a small upper 
limit on the width of the magnetic field distribution in the adjacent Ag.
We have also demonstrated that \bnmr\ can be used 
as a sensitive magnetic probe of 
spontaneous magnetic fields near an  interface.

We  thank R. Abasalti, D. Arseneau, S. Daviel, P. Dosanjh, and W. K. Park
for technical assistance. This work is supported by CIFAR, NSERC; 
and NRC through a contribution to TRIUMF. LHG acknowledges 
the support of the Office of Science, U.S. Department of
Energy under contracts DE- FG02-07ER46453 through the U. of Illinois
Frederick Seitz Materials Research Laboratory.



\end{document}